
\input phyzzx.tex
\def\kyoto{\centerline{\sl Department of Fundamental Sciences, FIHS}
          \centerline{\sl Kyoto University}
          \centerline{\sl Yoshida, Kyoto 606-01, Japan}}
\def\NP{Nucl.~Phys.~}
\def\PR{Phys.~Rev.~}

\def\PL{Phys.~Lett.~}
\VOFFSET = 1.2cm
\HOFFSET = .7cm
\pubnum{KUCP-63}
\date{February 1994}
\titlepage
\title{Instantons in Large Order of the Perturbative Series}
\author{Hideaki Aoyama
\foot{E-mail address: aoyama@shizen.yophy.kula.kyoto-u.ac.jp}}
\vskip 0.3cm
\kyoto
\vskip 1.5cm
\abstract
\doublespace
Behavior of the Euclidean path integral at large orders of the
perturbation series is studied.
When the model allows tunneling,
the path-integral functional in the zero instanton sector
is known to be dominated by bounce-like
configurations at large order of the perturbative series,
which causes non-convergence of the series.
We find that in addition to this bounce the perturbative functional has a
subleading peak at the
instanton and anti-instanton pair,
and its sum reproduces the non-perturbative
valley.

\endpage
\doublespace
\def\mysection#1{\vskip 1cm \noindent {\sl #1}}

\mysection{1. Introduction}

In quantum theories that allow tunneling phenomena
the perturbative series is known to diverge.\REFS
\brezin{F.~Br\'ezin,
   G.~Parisi and J.~Zinn-Justin \journal\PR &D16 (77) 408.}
\REFSCON\lipatov{L.~N.~Lipatov \journal Sov.~Phys.~JETP &45 (77) 216.}
\REFSCON\zinn{E.~Br\'ezin, J.-C.~Le Guillou and J.~Zinn-Justin \journal\PR
 &15 (77) 1544, 1558.}
\REFSCON\zinnbook{
{\sl \lq\lq Large-Order Behavior of Perturbation Theory"}
(edited by J.~C.~Le Guillou and J.~Zinn-Justin, North-Holland 1990),
and references contain herein.}
\refsend
Existence of instanton configuration for real coupling leads to
the result that perturbative coefficient behaves as $c_n \sim n^{n/2}$.
This leads to non-summable (not even Borel-summable) perturbative series.
Of course, this does not mean that such a theory is
ill-defined. In fact, one has to take into account the
``non-perturbative" effects.
However, these ``non-perturbative" effects
have some overlapping with the perturbative series at large order.
Therefore, it is of fundamental importance to seek a method
to do converging calculation, incorporating the perturbative as well as
non-perturbative contributions.
This kind of analysis becomes of practical interest when one is
faced with the problem of tunneling at higher energies,
such as the baryon and lepton number violation process
of the standard model at TeV energies.\REFS
\ring{A. Ringwald \journal \NP &B330 (90) 1.}
\REFSCON\esp{O. Espinosa  \journal \NP &B343 (90) 310.}
\REFSCON\allak{
H.~Aoyama and H.~Kikuchi \journal \PL &247B (90) 75
\journal \PR &43 (91) 1999
\journal {\sl Int.~J.~of Mod.~Phys.} &A7 (92) 2741.}
\refsend

In previous papers,\REFS\ao{
H.~Aoyama \journal{Mod.~Phys.~Lett.} &A7 (92) 1337.}
\REFSCON\aotam{H.~Aoyama and A.~M.~Tamra \journal \NP &B384 (92) 229.}
\refsend
the authors carried out analytical
studies of the perturbative series at large orders.
We used saddle-point approximation to the path-integral
functional at large orders to identify the dominating
configurations.
We found that bounce-like configurations
dominate and that configurations was responsible for
the diverging (non-Borel-summable) nature
of the perturbative series.  This was further
confirmed by various numerical calculations.

In this letter, we apply the above formalism to
the study of the contribution of the non-perturbative
configuration
to the perturbative functional.

\endpage
\mysection{2. Analysis}

\def\spg{S[\phi,g]}
For completeness we first go through the saddle point approximation
for the perturbative functional.
The actions of the models we deal with  are second order polynomials of the
coupling constant, which we denote by $g$.
$$S = c_0 - g c_1 + g^2 c_3 \eqno\eq$$
A one-dimensional quantum-mechanical model that fits in this
category is,
$$\spg = \int d\tau \left[
{1 \over 2}\left(  {\partial \phi \over \partial \tau} \right )^2
+ {1 \over 2} \phi^2 \left(1- g \phi \right)^2 \right] \ .  \eqn\action$$
The instanton solution corresponding to the tunneling from
$\phi = 0$ to $\phi = 1/g$ through the double-well potential is
known to have action $S^{\rm (I)} = 1/6g^2$.
This action leads to the WKB tunneling amplitude $e^{-1/6g^2}$.

\def\fn{F_n[\phi]}
\def\gabs{g_0}
In evaluating perturbative expansion of the partition function,
$${\cal Z} = \int {\cal D}\phi \ e^{-\spg}\ , \eqno\eq$$
in the zero-instanton sector,
we first expand integrand in powers of $g$;
$$ e^{-\spg} = \sum_{n=0}^\infty g^n \fn .\eqn\fndef$$
The perturbative functional $\fn$ can be expressed as a contour integral in the
complex $g$-plane,
$$\fn = {1 \over 2 \pi i} \ointop {d g \over g} e^{-(\spg+n\log g)}
\ . \eqno\eq$$
This can be evaluated using  the saddle point
approximation for the exponent $S+n\log g$
(which we denote by $\tilde S$ hereafter)
in the $n \rightarrow \infty$ limit.
There are always two saddle points.
Their positions depend on the behavior of the function $\phi(\tau)$:
In case $D \equiv 1 - (8n c_2 / c_1^2) > 0$, there are two saddle points on the
real axis of the $g$-plane, while otherwise there is a complex conjugate pair
of saddle points.
In the first case, we have found that we should choose
the contour to go through the
saddle-point of smaller absolute value.
In the latter case ($D<0$), the contour should be chosen to go through
{\it both} saddle points, which yields the following expression;
$$\fn = {2\over \sqrt{2 \pi}} \ {\rm Im}
 \left[ {1\over \sqrt{\tilde S_-^{\prime\prime} g_-^2}}
{e^{-S[\phi, g_-]} \over g_-^n} \right]\ , \eqn\fntwo$$
where we denoted the saddle point in the lower half-plane by $g_-$,
$$ g_- = \gabs e^{- i \theta}, \quad
\gabs \equiv \sqrt{n \over 2 c_2}, \quad
\cos\theta \equiv {c_1 \over \sqrt{8 n c_2}}. \eqn\gcomdef$$
In this case, since the contribution of the pair of saddle points
made  \fntwo\ different from the simple $e^{-S}$ form,
the equation that determines the dominating configuration is
different from ordinary equation of motion.
This allowed the dominating configuration to be a bounce
solution, which did not exist in the original theory.

\def\phii{\phi^{(I)}}

\mysection{3. Instantons in perturbative functional}

\def\pz{\phi_0}
\def\pii{\phi^{(I \bar I)}}
In order to look at the behavior of the perturbative functional,
we first numerically evaluate
$\fn$
in a two-dimensional subspace of the functional space of $\phi(\tau)$,
which contain both the bounce-like configuration  and the instanton pair.
We choose this space to be the space of two parameters $d$ and $\pz$,
which determine $\phi(\tau)$ as in the following;
$$\pii(\tau) = {\pz \over
	\left( 1 + e^{-\tau-d/2} \right)
	\left( 1 + e^{+\tau-d/2} \right)} \ . \eqn\anz$$
In \anz, if we choose $\pz = 1/g$ and $d$ large enough, we obtain
instanton and anti-instanton pair.  On the other hand, if we choose $d \sim 0$,
\anz\ is similar to bounce of height $\pz$.
[Since $\pii(\tau)$ is invariant under $(d, \pz) \rightarrow (-d, \pz e^d)$,
only $d > 0$ region shall be considered.]
For analysis, we have substituted \anz\ in \action\ and obtained the
expression  $S(d, \pz, g)$.
The exact integrand $e^{-S(d, \pz, g)}$ is plotted in
Fig.1.
\FIG\fullaction{
The plot of $e^{-S(d, \pz, 0.4)}$ in the $(d, \pz)$ space.
The axis $\phi=0$ is the classical minimum, where the
functional is equal to 1.
The ridge at $\pz=1/g$ for large $d$ corresponds to the pair of
instanton and anti-instanton with distance $d$.
The time-integration is actually done only for $\tau > 0$ region.
Therefore the height of the ridge is determined by {\it one}
instanton action $1/6g^2$.
}
The ridge at $\pz=1/g$ is the ``valley" (of action) in the valley methods.\REFS
\shuryak{E.V.Shuryak \journal \NP &B302 (88) 559,621.}
\REFSCON\balitsky{I.~I.~Balitzky and A.~V.~Yung \journal \PL &168B (86) 113;
\journal \NP \nextline &B274 (86) 475.}
\REFSCON\newvalley{H.~Aoyama and H.~Kikuchi \journal\NP &B369 (92) 219.}
\refsend
We have carried out
the $g$-expansion \fndef\ analytically (using Mathematica).
In Fig.2, we plot the resulting finite-order sum,
$${\cal F}_N(d, \pz) = \sum_{n=0}^{N} g^n F_n(d, \pz)\ . \eqn\fnn$$
\FIG\figtwo{Three dimensional
plots of \fnn, i.e., the integrand $e^{-S}$ cut-off at $N$-th
order. $N$ is chosen to be $2, 4, 6, 8$ for Figs.~(a), (b), (c), (d),
consequently.
The axis, scales, and the value of $g$ are the same as in Fig.1.
The plots are cut off from above at height $1$.}
As the order $N$ increases, the peak structure at $d \sim 0$ rises.
This corresponds to the fact that the perturbative functional $\fn$ is
dominated by bounce, as was shown in Re.~\ao\ and \aotam.
Furthermore, it is seen that the ridge structure (valley) is
reproduced at higher orders of the perturbation expansion.
More specifically, as the order increases, the
ridge structure extends to larger separation $d$.

\def\re#1{{\rm Re}(#1)}
Let us now carry out analytical study of what has been  observed above.
We take a pair of instanton and anti-instanton separated by a large
distance $d$ and approximate the value of
$\fnn$ as a function of $d$.
We first note that the configuration has
$$c_1 = {1\over g^3} (d - 3 + ...), \quad
c_2 = {1\over 2 g^4} (d + {11 \over 3} + ...) . \eqn\cc$$
[The omitted parts are non-leading and are irrelevant.]
We find that for $d < 8n g^2$, $D < 0$.
In estimating value of $\fn$ for that case, the most important factor is the
exponent $\re{S}$.
Using \cc and \gcomdef, we find that
$$\re{S} = S[\pii, g] + \delta S (g,d), \quad
\delta S (g,d) = {d\over 2 g^2} - {n \over 2} \  . \eqn\deltaresult$$
The contribution of the $g_-^n$ factor in \fntwo\ should also be looked
at, since potentially it might have large contribution.
The major part comes from the absolute value $g_0$, which is now
$$g_0 = g \sqrt{g^2 n \over d} , \eqn\gnresult$$
which adds logarithmic contributions to $\delta S$ in \deltaresult.
Combining \deltaresult\ and \gnresult, we find that
$$\eqalign{{\rm  Re}\left[ {e^{-S[\phi, g_-]} \over g_-^n} \right]
&= {e^{-S[\phii, g]} \over g^n} e^{-\Delta}, \crr
\Delta &= {n \over 2} \left( {d \over n g^2 } - 1
+ \log {n g^2 \over d}\right)
. \cr}
\eqn\deltafinal$$
Therefore
we find that in the $n-$th order perturbative polynomial $\fn$
has a peak at the instanton pair with distance
$$d = n g^2 + (\hbox{\rm non-leading terms}). \eqn\mainresult$$
Further,
since this pair has the ``right" weight, $e^{-1/3g^2}/g^n$,
as $g^n \fn$ is summed up, the valley is reproduced to larger and
larger $d$ with the right weight $e^{-1/3g^2}$.
This agrees with what can be observed in Fig.2.

\endpage
\mysection{4. Discussions}

In this letter we have seen that the perturbative functional integrand
$\fn$ have peaks at instanton and anti-instanton pair at distance
$d \sim n g^2$.  This implies the following:
Since the perturbative coefficient behaves as
$g^n n^{n/2}$ for $n \gg 1$, it starts to diverge
at around $n \sim 1/g^2$.
Therefore if there exists a way to
separate the ``perturbative" contribution from the
``non-perturbative" one ({\it i.e.} instanton pairs),
so that both give convergent results,
the effective cut-off of the perturbative series should be at $n \sim 1/g^2$.
On the other hand, according to \mainresult,
if the perturbation series is cut-off
at $n \sim 1/g^2$, it does not contain instanton pairs
of separation larger than O(1).
Since the width of instantons are of order one, for
$d > 1$ the instanton pair is well-separated.
Thus the cutoff at $n \sim 1/g^2$ serves to cutoff
the well-separated (and therefore definitely ``non-perturbative") instanton
and anti-instanton pairs.
[This is the advantage not shared by the method of fundamental region
discussed in Ref.\aotam].
Thus our result means that the cut-off
at $n \sim 1/g^2$ not only makes the perturbative
series well-behaved, but also separates the
well-defined non-perturbative valley.

The extention of the analysis to the field theories including the gauge
theories is straightforward.  The corresponding results will be published
in near future, together with the details of the analysis presented here.\Ref
\future{H.~Aoyama, Kyoto University preprint (1994), KUCP-66}

\endpage
\refout
\endpage
\figout
\bye